\documentclass[11pt]{article}
\usepackage{latexsym}
\usepackage{graphicx}
\usepackage{psfrag}
\newcommand{\be}{\begin{equation}}
\newcommand{\ee}{\end{equation}\noindent}
\newcommand{\bear}{\begin{eqnarray}}
\newcommand{\ear}{\end{eqnarray}\noindent}
\newcommand{\no}{\noindent}
\date{}
\renewcommand{\theequation}{\arabic{equation}}

\def\Eins{\mathord{1\hskip -1.5pt
\vrule width .5pt height 7.75pt depth -.2pt \hskip -1.2pt
\vrule width 2.5pt height .3pt depth -.05pt \hskip 1.5pt}}

\newcommand{\slD}{\raise.15ex\hbox{$/$}\kern-.57em\hbox{$D$}}
\newcommand{\slpartial}{\raise.15ex\hbox{$/$}\kern-.57em\hbox{$\partial$}}
\newcommand{\slG}{{{\dot G}\!\!\!\! \raise.15ex\hbox {/}}}

\def\eps{\epsilon}
\def\non{\nonumber}
\def\beqn*{\begin{eqnarray*}}
\def\eqn*{\end{eqnarray*}}

\def\square{\kern1pt\vbox{\hrule height 1.2pt\hbox{\vrule width 1.2pt
   \hskip 3pt\vbox{\vskip 6pt}\hskip 3pt\vrule width 0.6pt}
   \hrule height 0.6pt}\kern1pt}
\def\slash#1{#1\!\!\!\raise.15ex\hbox {/}}
\def\slashleft#1{#1\!\!\!\!\raise.15ex\hbox {/}}

\def\dps{\displaystyle}

\def\half{{1\over 2}}

\def\fourth{{1\over4}}

\def\eps{\epsilon}

\def\e{\mbox{e}}

\def\4piTD{{(4\pi T)}^{-{D\over 2}}}
\def\4piT4{{(4\pi T)}^{-2}}

\def\Tintm4{{\dps\int_{0}^{\infty}}{dT\over T}\,e^{-m^2T}
    {(4\pi T)}^{-2}}
\def\Tintm{{\dps\int_{0}^{\infty}}{dT\over T}\,e^{-m^2T}}

\def\tr{{\rm tr}\,}

\def\be{\begin{equation}}\def\ee{\end{equation}}
\def\bea{\begin{eqnarray}}\def\eea{\end{eqnarray}}
\def\ba{\begin{array}}\def\ea{\end{array}}
\def\bea{\begin{eqnarray}}\def\barr{\begin{array}}\def\earr{\end{array}}
\def\eea{\end{eqnarray}}
\begin{document} 
\newcommand{\ho}[1]{$\, ^{#1}$}
\newcommand{\hoch}[1]{$\, ^{#1}$}
\pagestyle{empty}
\renewcommand{\thefootnote}{\fnsymbol{footnote}}
\hfill {\sl BUCMP/02-03}

\hfill {\sl UMSNH-Phys/02-6}   
\vskip .4cm
\begin{center}
{\Large\bf Two-loop self-dual 
Euler-Heisenberg Lagrangians (I):}\\
\vspace{3.4pt}
{\Large\bf Real part and helicity amplitudes}
\vskip1.3cm
{\large Gerald V. Dunne}
\\[1.5ex]
{\it
Department of Physics\\
University of Connecticut\\
Storrs CT 06269, USA
}
\vspace{.8cm}

 {\large Christian Schubert
}
\\[1.5ex]
{\it
Instituto de F\'{\i}sica y Matem\'aticas
\\
Universidad Michoacana de San Nicol\'as de Hidalgo\\
Apdo. Postal 2-82\\
C.P. 58040, Morelia, Michoac\'an, M\'exico\\
schubert@itzel.ifm.umich.mx\\
}
\vspace{.2cm}
{\it
Center for Mathematical Physics, Mathematics Department\\
Boston University, Boston, MA 02215, USA}\\
\vspace{.2cm}
{\it
California Institute for Physics and Astrophysics\\
366 Cambridge Ave., Palo Alto, CA 94306, USA}

\vskip 2cm
 %
 {\large \bf Abstract}
\end{center}
\begin{quotation}
\noindent

We show that, for both scalar and spinor QED, the two-loop
Euler-Heisenberg effective Lagrangian for a constant Euclidean self-dual
background has an extremely simple closed-form expression in terms of the 
digamma function. Moreover, the scalar and spinor QED effective Lagrangians
are very similar to one another. These results are dramatic simplifications
compared to the results for other backgrounds.
We apply them to a calculation of the low energy limits
of the two-loop massive N-photon `all +'  helicity amplitudes. 
The simplicity of our results can be related to the
connection between self-duality, helicity and supersymmetry. 
\end{quotation}
\vskip 1cm
\clearpage
\renewcommand{\thefootnote}{\protect\arabic{footnote}}
\pagestyle{plain}

\section{Introduction: QED and QCD in constant fields}
\label{introduction}
\renewcommand{\theequation}{1.\arabic{equation}}
\setcounter{equation}{0}

The Euler-Heisenberg Lagrangian, one of the earliest results
in quantum electrodynamics \cite{eulhei,weisskopf}, describes the
complete one-loop amplitude involving a spinor loop
interacting non-perturbatively with a constant background
electromagnetic field. Euler and
Heisenberg found the following well-known integral representation 
for this effective Lagrangian:
\bear
{\cal L}_{\rm spin}^{(1)} 
&=&
-
{1\over 8\pi^2}
\int_0^{\infty}{dT\over T}
\,\e^{-m^2T}
\biggl[
{e^2ab\over \tanh(eaT)\tan(ebT)}
-{e^2\over 3} (a^2-b^2) -{1\over T^2}
\biggr]
\non\\
\label{L1spinren}
\ear
Here $T$ denotes the (Euclidean) propertime of the
loop fermion, and $a,b$ are related to the two
invariants of the Maxwell field by
$a^2-b^2=B^2-E^2, ab = {\bf E}\cdot{\bf B}$.
The subtractions of the terms of zeroeth and second order in $a,b$
corresponds to on-shell renormalization.

The Euler-Heisenberg Lagrangian
contains the information on the low -- enery limit
of the one--loop $N$--photon amplitudes
for any $N$. Moreover, it does so
in a form which is extremely convenient for the study
at low energies of nonlinear QED effects \cite{giesbook} such 
as photon--photon scattering \cite{eulhei},
photon dispersion \cite{toll,adler71}, and photon splitting
\cite{biabia,adler71,bamish,adlsch}.
For scalar QED, an analogous result was obtained by
Schwinger \cite{schwinger51}:
\bear
{\cal L}_{\rm scal}^{(1)} 
&=&
{1\over 16\pi^2}
\int_0^{\infty}{dT\over T}
\,\e^{-m^2T}
\biggl[
{e^2ab\over \sinh(eaT)\sin(ebT)}
+{e^2\over 6} (a^2-b^2) -{1\over T^2}
\biggr]
\non\\
\label{L1scalren}
\ear
The Lagrangians (\ref{L1spinren}) and (\ref{L1scalren}) are
real for a purely magnetic field, while
in the presence of an electric field there is
an absorptive part, indicating the process of
electron--positron (resp. scalar--antiscalar)
pair creation by the field \cite{schwinger51}.

The first radiative corrections to these Lagrangians,
describing the effect of an additional photon exchange in the loop, were
obtained in the seventies by Ritus \cite{ritusspin,ritusscal,ginzburg}. 
Using the exact propagators in a constant field found by
Fock \cite{fock} and Schwinger \cite{schwinger51}, and a proper-time
cutoff as the UV regulator, Ritus obtained the 
corresponding two-loop effective
Lagrangians ${\cal L}_{\rm scal,spin}^{(2)}$
in terms of certain two-parameter integrals.
Similar two-parameter integral
representations for ${\cal L}_{\rm scal,spin}^{(2)}$ where obtained
later by other authors, using either 
proper-time \cite{ditreuqed,rss}
or dimensional regularisation \cite{frss,korsch}.
Unfortunately, all of these double parameter integral representations
are quite complicated, 
so that it is much more difficult to study the weak- and
strong-field expansions at two-loops than at one-loop. 
Even more difficult becomes the analysis of the imaginary
part of the effective action \cite{lebrit}.
This is true even
for the special cases where the background is purely magnetic or purely
electric \cite{ritusscal,ginzburg}. 

However, the magnetic/electric cases are not the simplest ones 
which one can study in this context. As we will see in the following,
the constant field background which leads to maximal simplification
is the (Euclidean) self-dual one, given by
\footnote{This Euclidean self-duality condition should not be
confused with the similar condition $a=\pm b$ which {\sl can}
be realized in Minkowski space and could also be called
`self-duality'. See \cite{jgvlw} for a discussion of these
various dualities.}
\bear
F_{\mu\nu}=\tilde
F_{\mu\nu}\equiv \half\varepsilon_{\mu\nu\alpha\beta}F^{\alpha\beta}.
\label{defsd}
\ear
This is because for such a field the square of the
field strength tensor is proportional to the
Lorentz identity (see eq.(\ref{idF2}) below) 
so that the inevitable breaking of the
Lorentz invariance by the background field is in some sense
minimized. In Minkowski space the self-duality condition requires either E
or B to be complex. However, this does not imply that such backgrounds are
devoid of physical meaning. Rather, they should be interpreted in terms
of helicity projections \cite{dufish}. And indeed, it will be seen below
that it is quite straightforward to extract, from the effective action
for such a self-dual field, the low energy limit of the two-loop amplitude
for the  scattering of N photons with all equal helicities. In addition,
we see at least four more good reasons for studying this particular
self-dual case in detail:

First, a 
detailed comparison at the one-loop level shows that
the self-dual case with real $\bf B$ and complex $\bf E$
(corresponding to a real self-dual field strength $f$, where $f^2=\fourth
F_{\mu\nu}F^{\mu\nu}$, and called the `magnetic' case in the following) 
leads to an effective Lagrangian whose properties are only marginally
different from the ones of the magnetic Euler-Heisenberg Lagrangian,
while the case with a real electric and complex magnetic field
(corresponding to $f$ imaginary, and called the `electric' case in the
following) is a very good analogue
of the electric case. This leads us to expect
that the study of the self-dual case at higher loops
may yield useful information on the generic properties
of Euler-Heisenberg Lagrangians.

Second, many quantities of physical interest in quantum
electrodynamics are computable in Euclidean space. This includes
the renormalization constants, and in particular
the QED $\beta$--function.

Third, the constant self-dual Euclidean backgrounds that we consider 
here for QED generalize in a very simple way to the case of quasi-abelian
self-dual constant backgrounds in QCD. Such backgrounds have been 
studied extensively in QCD as they have the special property \cite{leutwyler}
that among the covariantly constant gluon backgrounds, only the self-dual
quasi-abelian background is stable (at one-loop) under fluctuations. This has
led to extensive studies of quark and gluon loops in such a background 
\cite{leutwyler,finjord,flory,elizalde}. 
Also at one-loop, much is known about QCD
in the presence of arbitrary covariantly constant background fields. One-loop
results along the lines of the Euler-Heisenberg formulas (\ref{L1spinren}) 
and (\ref{L1scalren}) can be found in \cite{dufframon,bamasa}. 
The purely chromomagnetic covariantly constant background is unstable 
\cite{nielsen,pagels,leutwyler}. The renormalization of these one-loop 
results in pure QCD is complicated by infrared problems, problems
which become even worse at the two-loop level \cite{sascza}. 

A fourth motivation for studying constant quasi-abelian self-dual backgrounds 
is that
they could provide useful information about effective Lagrangians in 
other self-dual backgrounds, in particular instantons
\cite{thooft,carlitz,kwon}.

As our main result in this  paper, we will 
show that in the self-dual case, at two-loop, {\sl all} 
parameter integrals can be done in closed form, yielding the following
simple expressions for the two-loop scalar and spinor 
QED effective Lagrangians:
\bear
{\cal L}_{\rm scal}^{(2)(SD)}(\kappa)
&=&
\alpha \,{m^4\over (4\pi)^3}\frac{1}{\kappa^2}\left[
{3\over 2}\xi^2 (\kappa)
-\xi'(\kappa)\right]
\label{2lscintro}\\
{\cal L}_{\rm spin}^{(2)(SD)}(\kappa)
&=&
-2\alpha \,{m^4\over (4\pi)^3}\frac{1}{\kappa^2}\left[
3\xi^2 (\kappa)
-\xi'(\kappa)\right]
\label{2lspintro}
\ear
Here we have defined the convenient dimensionless parameter
\begin{eqnarray}
\kappa\equiv \frac{m^2}{2e\sqrt{f^2}}
\label{kappa}
\end{eqnarray}
where $f^2=\fourth F_{\mu\nu}F^{\mu\nu}$, 
as well as the important function
\bear
\xi(x)\equiv -x\Bigl(\psi(x)-\ln(x)+{1\over 2x}\Bigr)
\label{defxi}
\ear
with $\psi$ being the digamma function
$\psi(x)=\Gamma^\prime(x)/\Gamma(x)$. The scalar QED formula 
(\ref{2lscintro}) was already presented in \cite{dunsch2}.

The extreme simplicity of these results, and the obvious similarity between 
the two-loop effective Lagrangians in the scalar and spinor cases, provides
a new motivation for studying self-dual backgrounds. Our results provide
a new example of the well-known connections among self-duality, helicity
and supersymmetry. Self-dual fields have definite helicity \cite{dufish},
and are closely related to supersymmetry \cite{thooft,dadda,brolee}. 
One consequence of this connection is that there exist remarkably simple
formulas for loop amplitudes when the external fields have all (or almost all)
helicities being equal. Such simplifications 
have been known for a long time in massless QCD at the tree level 
\cite{ttwu}. More recently they have been 
studied extensively at the one- and two loop level, in abelian
\cite{mahlon,bddgw,bgmv} as well as in nonabelian 
\cite{mangano,bernreview,bededi}
gauge theory. 
The close interplay been self-duality, supersymmetry and
integrability has also been explored in this context. 
The simplicity of the structure of the 
QED/QCD amplitudes with all (almost all)
helicities alike is thought to be deeply related 
to the integrability properties of self-dual Yang-Mills fields
\cite{bardeen,daniel}.
In the present paper the connection between self-duality and helicity
will be used to obtain, from the above results for the
self-dual two-loop
effective Lagrangians, the low energy limits of the QED
$N$ - photon amplitudes with all helicities alike.

The outline of the paper is as follows. In section 2 we write down
the one-loop self-dual
Euler-Heisenberg Lagrangians for scalar and spinor QED, 
as well as for the `quasi-abelian' special case in QCD.
In section 3 we compute the two-loop corrections to these
Lagrangians for scalar and spinor QED,
using the `string-inspired' worldline formalism 
\cite{berkos,strassler} along the lines of \cite{ss1,ss3,rss,frss}. 
This formalism is based on the representation of effective
actions in terms of first-quantized particle path integrals,
and has turned out to be highly convenient for computations
involving constant external fields 
\cite{ss1,mckshe,cadhdu,gussho,shaisultanov,rss,adlsch,frss}
(see \cite{review} for a review).
In section 4 we comment on the special properties
of self-dual fields with respect to helicity and supersymmetry,
and we use some of these properties to explain the similarity
of our two-loop scalar and spinor QED results.
We then proceed in section 5 to an explicit
calculation of the low energy limits of the massive
all `+' helicity N photon amplitudes, at one and two loops,
in scalar and spinor QED.
Section 6 contains a summary, and some possible future
directions of work. In an accompanying paper \cite{sdtwo} we study the
weak- and strong-field expansions of these two-loop effective
Lagrangians, and use them to test the techniques of Borel summation, and
the associated analytic continuation properties and structure of the
imaginary part of the two-loop self-dual Euler-Heisenberg Lagrangians for
the `electric' case.

\section{One loop self--dual Euler-Heisenberg Lagrangians}
\label{section1lsd}
\renewcommand{\theequation}{2.\arabic{equation}}
\setcounter{equation}{0}

\subsection{Scalar QED}

For scalar QED, the (unrenormalized) one-loop effective
Lagrangian in a constant $F_{\mu\nu}$ background is \cite{review}
\begin{eqnarray}
{\cal L}^{(1)}_{\rm scal}
(F)&=&
{(4\pi )}^{-{D\over 2}}
\int_0^{\infty}{dT\over T}e^{-m^2T}T^{-{D\over 2}} 
{\rm det}^{-{1\over 2}}
\biggl[{\sin(eFT)\over {eFT}}
\biggr]
\label{L1scalgen}
\end{eqnarray}
\noindent
Here $T$ is the proper-time variable for the loop scalar, and $D$ is the
spacetime dimension. We would like to evaluate this integral for the case
of a self-dual field, 
$F_{\mu\nu}=\tilde F_{\mu\nu}\equiv\half\varepsilon_{\mu\nu\alpha\beta}
F^{\alpha\beta}$. The self-duality condition implies that
\bear
F^2 &=& -f^2\Eins
\label{idF2}
\ear\no
where $f^2=\fourth F_{\mu\nu}F^{\mu\nu}$, 
and $\Eins$ denotes
the identity matrix in Lorentz space. Therefore
\bear
{\rm det}^{-{1\over 2}}
\biggl[{\sin(eFT)\over {eFT}}
\biggr]
&=&
\Bigl({efT\over\sinh(efT)}\Bigr)^{D\over 2}
\label{detselfdual1loop}
\ear
Renormalization involves subtracting the free
field contribution, and also a charge renormalization,
corresponding to subtracting the logarithmically divergent $O(f^2)$
term. This leads to the
following renormalized one-loop effective Lagrangian:
\bear
{\cal L}_{\rm scal}^{(1)(SD)}(\kappa)
=
{m^4\over (4\pi)^2}\frac{1}{4\kappa^2}
\int_0^{\infty}
{dZ\over Z}\e^{-2\kappa Z}
\biggl[
{1\over \sinh^2(Z)}-\frac{1}{Z^2}+{1\over 3}
\biggr]
\label{1lsca}
\ear
We remark that
this integral can be expressed in terms of the
function
$\ln\Gamma(x)$ and its integral function:
\bear
{\cal L}_{\rm scal}^{(1)(SD)}(\kappa) 
&=& {m^4\over (4\pi)^2}\frac{1}{\kappa^2}\left[-{1\over
12}\ln(\kappa) +\zeta'(-1)+\Xi(\kappa)\right]
\label{1lsc}
\ear 
where the function $\Xi(x)$ is defined as follows:
\bear
\Xi(x)\equiv \int_0^x dy\,\ln\Gamma(y)
-x\ln\Gamma(x)+{x^2\over 2}\ln(x)-{x^2\over 4}-{x\over 2}
\label{defXi}
\ear
This one-loop result (\ref{1lsc}) may equivalently be expressed in terms
of the Hurwitz zeta function (see, e.g., \cite{ditreuqed,matt}), but we
prefer to use this representation here, since 
the two-loop result is a simple function of 
$\xi(\kappa) =\Xi^\prime(\kappa)$. 

\subsection{Spinor QED}

For spinor QED, the unrenormalized one-loop effective Lagrangian in a
constant $F_{\mu\nu}$ background is
\begin{eqnarray}
{\cal L}^{(1)}_{\rm spin}
(F)=-2
{(4\pi )}^{-{D\over 2}}
\int_0^{\infty}{dT\over T}\, e^{-m^2T}T^{-{D\over 2}} 
{\rm det}^{-{1\over 2}}
\biggl[{\tan(eFT)\over {eFT}}
\biggr]
\label{L1spingen}
\end{eqnarray}
\noindent
For a self-dual field 

\bear
{\rm det}^{-{1\over 2}}
\biggl[{\tan(eFT)\over {eFT}}
\biggr]
=
\Bigl({efT\over\tanh(efT)}\Bigr)^{D\over 2}
\label{detsd1loopspin}
\ear
After renormalization, this leads to the following effective Lagrangian:

\bear
{\cal L}_{\rm spin}^{(1)(SD)}(\kappa)
=
-{m^4\over (4\pi)^2}\frac{1}{2\kappa^2}
\int_0^{\infty}
{dZ\over Z}\,\e^{-2\kappa Z}
\biggl[
{\coth^2 Z}-\frac{1}{Z^2}-\frac{2}{3}
\biggr]
\label{1lspa}
\ear

\subsection{QCD}

In the nonabelian case, the natural notion of a
constant field strength is covariant constancy, i.e.
${\cal D}_{\alpha}^{ab}F^b_{\mu\nu}=0$. This does  not imply
that $F^b_{\mu\nu}$ is a constant matrix, so that further
assumptions need to be made to arrive at Euler-Heisenberg type
formulas. The simplest such case is the `quasi-abelian' one,
where $A_{\mu}=A^a_{\mu}T^a$, and therefore also 
$F_{\mu\nu}$, are assumed to point in a fixed
direction in colour space:

\bear
A_{\mu}^a(x) &=& n^a A_{\mu}(x), \qquad
F_{\mu\nu}^a(x) = n^a F_{\mu\nu}(x)
\label{defquasiabelian}
\ear
The computation of the
one-loop effective action with such a field is identical
to the abelian one for the scalar and spinor loop cases,
since the colour degree of freedom manifests itself only
in a global colour trace. Thus for the self-dual case
one obtains (compare (\ref{L1scalgen}),(\ref{L1spingen}))

\begin{eqnarray}
{\cal L}^{(1)}_{\rm scal}
(F)&=&
{\rm tr}_c
\int_0^{\infty}{dT\over T}\,e^{-m^2T}(4\pi T)^{-{D\over 2}} 
\Bigl({gfT{\slash n}\over\sinh(gfT{\slash n})}\Bigr)^{D\over 2}
\label{L1scalqcd}\\
{\cal L}^{(1)}_{\rm spin}
(F)&=&
-2
{\rm tr}_c
\int_0^{\infty}{dT\over T}\,e^{-m^2T}(4\pi T)^{-{D\over 2}} 
\Bigl({gfT{\slash n}\over\tanh(gfT{\slash n})}\Bigr)^{D\over 2}
\label{L1spinqcd}
\end{eqnarray}
For the gluon loop one finds, in this quasi-abelian constant field
case, the result \cite{bamasa,matt,rss}
\bear
{\cal L}^{(1)}_{\rm glu}
(F)&=&
\half
{\rm tr}_c
\int_0^{\infty}{dT\over T}(4\pi T)^{-{D\over 2}} 
\exp\biggl\lbrack
-\half\tr_L
\ln \Bigl(\frac{\sin(gFT)}{gFT}\Bigr)
\biggr\rbrack
\tr_L \cos (2gFT)
\non\\
\label{1lqcd}
\ear
In these formulas, $\tr_L$ denotes the Lorentz trace, $\tr_c$ the
colour trace, and we defined ${\slash n}\equiv n^aT^a$.
For the self-dual case, (\ref{1lqcd}) can be written as
\bear
{\cal L}^{(1)}_{\rm glu}
(F)&=&
2{\rm tr}_c
\int_0^{\infty}{dT\over T}(4\pi T)^{-{D\over 2}} 
\Bigl({gfT{\slash n}\over\sinh(gfT{\slash n})}\Bigr)^{D\over 2}
\Bigl(1+2\sinh^2(gfT{\slash n})\Bigr)
\non\\
\label{1lqcdsd}
\ear
Let us specialize further to the case where all
particles are in the adjoint representation and massless.
We can then combine the contribution of $m$ (real) scalars, $n$ (Weyl)
spinors, the gauge boson and its ghost (which gives minus the contribution
of a complex scalar) into (putting $D=4$)
\bear
{\cal L}^{(1)}_{\rm total}
(F)
&=&
\half\tr_c
\int_0^{\infty}{dT\over T}(4\pi T)^{-2} 
\biggl\lbrace
(8-2n)\bigl(gfT{\slash n}\bigr)^2
+(m-2n+2)
\Bigl[\frac{gfT{\slash n}}{\sinh(gfT{\slash n})}
\Bigr]^2
\biggr\rbrace
\non\\
\label{1lqcdtotal}
\ear
For $m=6$ and $n=4$ we recover the well-known fact 
that the one-loop self-dual effective action vanishes for $N=4$ SYM theory
\cite{dadda,dufish,fratsesusy}.

\section{Two loop self--dual Euler-Heisenberg Lagrangians}
\label{section2lsd}
\renewcommand{\theequation}{3.\arabic{equation}}
\setcounter{equation}{0}

\subsection{Scalar QED}

In \cite{rss,frss} the two-loop Euler-Heisenberg Lagrangian
in (Euclidean) Scalar QED was obtained in terms of the
following fourfold parameter integral,
\begin{eqnarray}
{\cal L}^{(2)}_{\rm scal}
(F)&=&
{(4\pi )}^{-D}
\Bigl(-{e^2\over 2}\Bigr)
\int_0^{\infty}{dT\over T}e^{-m^2T}T^{-{D\over 2}} 
\int_0^{\infty}d\bar T 
\int_0^T d\tau_a
\int_0^T d\tau_b
\nonumber\\
&\phantom{=}&\times
{\rm det}^{-{1\over 2}}
\biggl[{\sin(eFT)\over {eFT}}
\biggr]
{\rm det}^{-{1\over 2}}
\biggl[
\bar T 
-{1\over 2}
{\cal C}_{ab}
\biggr]
\langle
\dot y_a\cdot\dot y_b\rangle
\label{Gamma2scal}
\end{eqnarray}
\noindent
Here $T$ and $\bar T$ represent the scalar
and photon proper-times, and $\tau_{a,b}$ the
endpoints of the photon insertion moving around
the scalar loop. 
${\cal C}_{ab}$ and 
$\langle\dot y_a\cdot\dot y_b\rangle$
are given by
\bear
{\cal C}_{ab}&=& 
{\cal G}_B(\tau_a,\tau_a)
-{\cal G}_B(\tau_a,\tau_b)
-{\cal G}_B(\tau_b,\tau_a)
+{\cal G}_B(\tau_b,\tau_b)
\nonumber\\
\langle
\dot y_a\cdot\dot y_b\rangle
&=&
{\rm tr}
\biggl[
\ddot{\cal G}_{Bab}+{1\over 2}
{(\dot {\cal G}_{Baa}-\dot {\cal G}_{Bab})
(\dot {\cal G}_{Bab}-\dot {\cal G}_{Bbb})
\over
{\bar T -{1\over 2}{\cal C}_{ab}}}
\biggr]
\label{defCabWick}
\ear
\noindent
They are expressed in terms of the worldline Green's
function ${\cal G}_B$ and its first and second derivatives
\cite{rss}:
\bear
{\cal G}_{B}(\tau_1,\tau_2) &=&
{1\over 2{(eF)}^2}\biggl({eF\over{{\rm sin}(eFT)}}
{\rm e}^{-ieFT\dot G_{B12}}
+ieF\dot G_{B12} -{1\over T}\biggr)
\non\\
\dot{\cal G}_B(\tau_1,\tau_2)
&=&
{i\over eF}\biggl({eF\over{{\rm sin}(eFT)}}
{\rm e}^{-ieFT\dot G_{B12}}-{1\over T}\biggr)
\nonumber\\
\ddot{\cal G}_{B}(\tau_1,\tau_2)
&=& 2\delta (\tau_1 -\tau_2) -2{eF\over{{\rm sin}(eFT)}}
{\rm e}^{-ieFT\dot G_{B12}}
\label{calGBetc}
\end{eqnarray}
\noindent
These are Lorentz matrices, and the above formulas should be
understood as power series in the field strength tensor $F_{\mu\nu}$.
The scalar function $\dot{G}_{B12}$ is given by $\dot{G}_{B12}={\rm sign}(\tau_1-\tau_2)-2(\tau_1-\tau_2)/T$. 
Due to the translation invariance of those Green's functions
one of the integrations $\int d\tau_{a,b}$ is redundant, so that
we will set $\tau_b =0$ in the following.

Removing the second derivative $\ddot {\cal G}_B$ by a
partial integration with respect to $\tau_a$ or $\tau_b$
one can obtain the equivalent integral

\begin{eqnarray}
{\cal L}^{(2)}_{\rm scal}(F)&=&
{(4\pi )}^{-D}
\Bigl(-{e^2\over 2}\Bigr)
\int_0^{\infty}{dT\over T}e^{-m^2T}T^{-{D\over 2}} 
\int_0^{\infty}d\bar T 
\int_0^T d\tau_a
\int_0^T d\tau_b
\nonumber\\
&\phantom{=}&\times\,
{\rm det}^{-{1\over 2}}
\biggl\lbrack{\sin(eFT)\over {eFT}}
\biggr\rbrack
{\rm det}^{-{1\over 2}}
\biggl[
\bar T -{1\over 2}
{\cal C}_{ab}
\biggr\rbrack\nonumber\\
&\phantom{=}&\times
{1\over 2}
\Biggl\lbrace
{\rm tr}\dot{\cal G}_{Bab}
{\rm tr}
\biggl\lbrack
{\dot{\cal G}_{Bab}
\over
{\bar T -{1\over 2}{\cal C}_{ab}}}
\biggr\rbrack
+{\rm tr}
\biggl[
{(\dot {\cal G}_{Baa}-\dot {\cal G}_{Bab})
(\dot {\cal G}_{Bab}-\dot {\cal G}_{Bbb})
\over
{\bar T -{1\over 2}{\cal C}_{ab}}}
\biggr]
\Biggr\rbrace
\label{Gamma2scalpI}
\end{eqnarray}
\noindent
We would again like to evaluate this integral for the case of a
self-dual field. The worldline correlators
(\ref{calGBetc}) for such a field simplify to the following \cite{vv}:

\bear
{\cal G}_{B12} &=&
{T\over 2}
\biggl[
{1\over Z^2}-{\cosh(Z\dot G_{B12})\over Z\sinh(Z)}
\biggr]
\Eins + 
i{T\over 2Z^2}\Bigl[{\sinh(Z\dot G_{B12})\over\sinh(Z)}
-\dot G_{B12}\Bigr]{\cal Z}\non\\
\dot{\cal G}_{B12} &=&
{\sinh(Z\dot G_{B12})\over\sinh(Z)}\Eins
-i\Bigl[{\cosh(Z\dot G_{B12})\over Z\sinh(Z)}-{1\over Z^2}\Bigr]
{\cal Z}
\non\\
\ddot {\cal G}_{B12} &=&
\Bigl[
2\delta_{12} -{2\over T}{Z\cosh(Z\dot G_{B12})\over\sinh(Z)}
\Bigr]\Eins
+i{2\over T}{\sinh(Z\dot G_{B12})\over\sinh(Z)}
{\cal Z}
\label{calGBetcselfdual}
\ear\no
where ${\cal Z}_{\mu\nu}\equiv eTF_{\mu\nu}$. 
The first determinant factor is already known from
the one-loop calculation, eq. (\ref{detselfdual1loop}).
The Lorentz matrix ${\cal C}_{ab}$, being an even function of $F$,
for this background becomes scalar:
\bear
{\cal C}_{ab} &=& -C_{ab} \Eins, \non\\
C_{ab} &=& T{\cosh(Z)-\cosh(Z\dot G_{Bab})\over
Z\sinh(Z)}
\label{Cabselfdual}
\ear\no
Therefore in the self-dual case the $\bar T$ integrals in 
eqs.(\ref{Gamma2scal}),(\ref{Gamma2scalpI}) can be done trivially.
In the same way as in the magnetic field case \cite{rss,frss}, 
further simplification is achieved 
by taking the following linear combination
of the representations (\ref{Gamma2scal}) and (\ref{Gamma2scalpI}),
\begin{equation}
{\cal L}^{(2)(SD)}_{\rm scal}(F)=
{{D-1}\over D}\times
{\rm eq.}
(\ref{Gamma2scal})
+{1\over D}\times
{\rm eq.}
(\ref{Gamma2scalpI})
\label{Gamma2scaloptim}
\end{equation}
\noindent
After the usual rescaling to the unit circle,
$\tau_{a,b}=Tu_{a,b}$,
and setting $u_b=0, u_a=u$, we end up with the
following twofold integral,
\footnote{Note that the $\delta_{ab}$ - term contained in
$\ddot{\cal G}_{Bab}$ has been deleted, since here the
photon proper-time integral is a
tadpole type integral 
which vanishes in dimensional regularization.} 
\bear
{\cal L}^{(2)(SD)}_{\rm scal}(f)
=
{e^2\over 2(4\pi)^D}
\int_0^{\infty}{dT\over T}\,\e^{-m^2T}
T^{2-D}{Z^D\over\sinh^2(Z)}
2^{D\over 2}
\Bigl\lbrace I_1(Z,D) + 2{D-1\over D-2} I_2(Z,D) \Bigr\rbrace
\label{Gamma2scalfinalint}
\ear\no
where
\bear
I_1(Z,D) &=& \int_0^1du
{1\over [\cosh(Z)-\cosh(Z(1-2u))]^{\epsilon\over 2}}
\non\\
I_2(Z,D) &=& \int_0^1du
{\cosh(Z(1-2u))\over [\cosh(Z)-\cosh(Z(1-2u))]^{1+{\epsilon\over 2}}}
\label{I1,I2}
\ear\no
($\epsilon = D-4$).
We are only interested in terms of order $O(F^4)$ and higher in this
Lagrangian, since the $O(F^2)$ term contributes only to
photon wave function renormalization. For those terms the 
final $T$ - integration
is finite, so that in $I_1$ we can set $\epsilon =0$ 
immediately. The same is not possible for $I_2$, since this
integral for $D=4$ has a divergence at the points where the
two photon end points become coincident, $u=0,1$.
For the calculation of this integral we split it in the following way,

\bear
I_2(Z,D) &=& 
\int_0^1 du \;
\Bigl[\cosh(Z(1-2u))-\cosh\Bigl((1-{\eps\over 2})Z(1-2u)\Bigr)\Bigr]
\non\\&&\hspace{-30pt}\times
\biggl[{1\over [\cosh(Z)-\cosh(Z(1-2u))]^{1+{\epsilon\over 2}}}
-{1\over [2Z\sinh(Z)u]^{1+{\epsilon\over 2}}}
-{1\over [2Z\sinh(Z)(1-u)]^{1+{\epsilon\over 2}}}
\biggr]
\non\\
&& +
\int_0^1du\;
{\cosh(Z(1-2u))-\cosh\Bigl((1-{\eps\over 2})Z(1-2u)\Bigr)
\over
[2Z\sinh(Z)u]^{1+{\epsilon\over 2}}
}
\non\\
&& +
\int_0^1du\;
{\cosh(Z(1-2u))-\cosh\Bigl((1-{\eps\over 2})Z(1-2u)\Bigr)
\over
[2Z\sinh(Z)(1-u)]^{1+{\epsilon\over 2}}
}
\non\\&&
+ 
\int_0^1du\;
{\cosh\Bigl((1-{\eps\over 2})Z(1-2u)\Bigr)
\over [\cosh(Z)-\cosh(Z(1-2u))]^{1+{\epsilon\over 2}}}
\non\\
\label{splitI2}
\ear\no
Of these four integrals the first one is of order O$(\epsilon)$
and can be omitted. The second and third ones are equal
and  elementary,

\bear
\int_0^1du\;
{\cosh(Z(1-2u))-\cosh\Bigl((1-{\eps\over 2})Z(1-2u)\Bigr)
\over
[2Z\sinh(Z)u]^{1+{\epsilon\over 2}}
}
&=&
- \half
+ O(\eps)
\non\\
\label{secondint}    
\ear
The fourth integral can be calculated by transforming to the
variable $\phi = (1-2u)Z$, and using 
formula 22.10.11 of \cite{abramowitz}
(see \cite{szego} for a general technique for computing this
type of integral).
The result is

\bear
{1\over Z} \int_0^Z d\phi\, 
{\cosh\bigl((1-{\epsilon\over 2})\phi\bigr)
\over [\cosh(Z)-\cosh(\phi)]^{1+{\epsilon\over 2}}}
&=& 2^{-1-{\epsilon\over 2}}
B(-{\epsilon\over 2},-{\epsilon\over 2})
{\cosh(Z)\over Z\Bigl(\sinh(Z)\Bigr)^{1+\epsilon}}
\label{thirdint}
\ear
with $B$ the Euler Beta function.
Expanding in $\epsilon$ we find

\bear
{\cal L}^{(2)(SD)}_{\rm scal}(f_0)&=&
{e_0^2\over (4\pi)^4}
\int_0^{\infty}{dT\over T^3}
\,\e^{-m_0^2T}
\Biggl\lbrace
\cosh(Z)\Biggl({Z\over\sinh(Z)}\Biggr)^3
\biggl\lbrack -{12\over\eps} + 2
+ 12\ln (4\pi T)
\non\\
&&\hspace{120pt} 
- 12\ln\Bigl({Z\over\sinh(Z)}\Bigr)\biggr\rbrack
- 4{Z^4\over\sinh^2(Z)}\Biggr\rbrace
\label{Lreg}
\ear
We have written this result for the Lagrangian
in terms of bare quantities, since it still
requires renormalization.
Mass renormalization requires us
to subtract a term 
$\delta m_0^2 {{\partial}\over{\partial} m_0^2}
{\cal L}^{(1)(SD)}_{\rm scal}(f_0)$, 
where ${\cal L}^{(1)(SD)}_{\rm scal}(f_0)$ is the one-loop
Euler-Heisenberg Lagrangian (\ref{1lsca}), 
and $\delta m_0^2$ 
the one-loop mass shift, both calculated in dimensional
regularization. The mass shift is

\bear
\delta m_0^2 &=&
m_0^2
{\alpha_0\over 4\pi}
\Bigl[
-{6\over\epsilon}
+7
-3
\bigl[\gamma - \ln (4\pi)\bigr]
-3\ln (m_0^2)
\Bigr]+O(\epsilon)
\, 
\label{deltamscaldim}
\ear
Noting that 

\bear
{\cosh(Z)\over\sinh^3(Z)} &=& -\half {d\over dZ}\, {1\over\sinh^2(Z)}
\label{partint}
\ear
we perform a partial integration on the first term on the
right hand side of (\ref{Lreg}). After this the pole term in it
takes the same form as the pole of the mass shift term above.
Subtracting the mass shift term we end up with the following expression
for the renormalized Lagrangian,
\bear
{\cal L}^{(2)(SD)}_{\rm scal}(f)&=& 
{e^2\over(4\pi)^4}
\int_0^{\infty}{dT\over T^3}\,\e^{-m^2T}
\Bigl({Z\over\sinh(Z)}\Bigr)^2
\biggl\lbrace
-4Z^2
+3m^2T\Bigl[1-\gamma +\ln\Bigl({Z\over\sinh(Z)m^2T}\Bigr)
\Bigr]
\biggr\rbrace
\non\\
\label{Lfinal}
\ear
This can also be written as
\bear
{\cal L}^{(2)(SD)}_{\rm scal}(\kappa)= 
m^4 {e^2\over(4\pi)^4} \frac{1}{\kappa^2}
\int_0^{\infty}{dZ\over Z^3}\,\e^{-2\kappa Z}
\Bigl({Z\over\sinh(Z)}\Bigr)^2
\biggl\lbrace
-Z^2
+\frac{3}{2}\kappa Z\Bigl[1-\gamma +\ln\Bigl({Z\over\sinh(Z)}\Bigr)
- \ln(2\kappa Z) 
\Bigr]
\biggr\rbrace
\non\\
\label{Lfinalscaled}
\ear
Note that, since we have not yet performed the photon wave
function renormalization, this formula is correct
only beginning from $O(\frac{1}{\kappa^4})$.

For the calculation of this integral, we go back to the
regularized version, eqs. (\ref{Gamma2scalfinalint}) -- (\ref{thirdint}),
and note that it involves essentially only the integral

\bear
\int_0^{\infty}
dx\,
\e^{-2\kappa x}
\sinh^{\gamma}(x)
&=&
2^{-\gamma - 1}B\Bigl(\gamma +1,\kappa-{\gamma\over 2}\Bigr)
\label{mountainintegral}
\ear
However, to be able to take the limit $\epsilon\to 0$ we instead
introduce

\bear
I(\kappa,\epsilon) &\equiv&
\int_0^{\infty}
dx\,\e^{-2\kappa x}
\biggl\lbrack
\frac{1}{\sinh^{2+\epsilon}(x)}
-\frac{1}{x^{2+\epsilon}}
\biggr\rbrack
\label{defI}
\ear
Including the mass renormalization term, and
again disregarding $O(f^2)$ terms, we can then write

\bear
{\cal L}^{(2)(SD)}_{\rm scal}(f_0)
- 
\delta m_0^2 {{\partial}\over{\partial} m_0^2}
{\cal L}^{(1)(SD)}_{\rm scal}(f_0)
&=&
m^4 {e^2\over(4\pi)^4} \frac{1}{2\kappa}
\biggl\lbrace
3\bigl(1-\gamma-\ln(2\kappa)\bigr)
+{1\over\kappa}{\partial\over\partial\kappa}
+3{\partial\over\partial\epsilon}
\biggr\rbrace
I(\kappa,\epsilon)\!\!\mid_{\epsilon=0}
\non\\
\label{L2scalbyI}
\ear

Using (\ref{mountainintegral}) this yields, after an easy
computation, our final result (\ref{2lscintro}),

\bear
{\cal L}_{\rm scal}^{(2)(SD)}(\kappa)
&=&
\alpha \,{m^4\over (4\pi)^3}\frac{1}{\kappa^2}\left[
{3\over 2}\xi^2 (\kappa)
-\xi'(\kappa)\right]
\nonumber
\ear
where we have finally subtracted out the terms of
order $O(f^0)$, $O(f^2)$.

\subsection{Spinor QED}
\label{section2lspin}

The calculations for the spinor QED case 
are very similar, so we content ourselves
primarily with a presentation of the results.

For the two-loop Euler-Heisenberg Lagrangian in Euclidean
spinor QED, the worldline formalism 
leads to the following parameter integral,
which is completely analogous to  
(\ref{Gamma2scalpI})
\cite{rss,frss}:

\begin{eqnarray}
{\cal L}^{(2)}_{\rm spin}
(F)&=&
{(4\pi )}^{-D}
e^2
\int_0^{\infty}{dT\over T}e^{-m^2T}T^{-{D\over 2}} 
\int_0^{\infty}d\bar T 
\int_0^T d\tau_a
\int_0^T d\tau_b
\nonumber\\
&\phantom{=}&\times
{\rm det}^{-{1\over 2}}
\biggl[{\tan(eFT)\over {eFT}}\left(\bar{T}-\frac{1}{2}{\cal C}_{ab}\right)
\biggr]\frac{1}{2}\biggl\{ \tr \dot{\cal G}_{Bab}\,\tr\biggl[
\frac{\dot{\cal G}_{Bab}}{\bar T -{1\over 2}{\cal
C}_{ab}}\biggr]-\tr {\cal G}_{Fab}\,\tr\biggl[
\frac{{\cal G}_{Fab}}{\bar T -{1\over 2}{\cal
C}_{ab}}\biggr] \nonumber\\ 
&\phantom{=}&+\tr\biggl[{(\dot{\cal G}_{Baa}-\dot{\cal
G}_{Bab})(\dot{\cal G}_{Bab}-
\dot{\cal G}_{Bbb}+ {\cal G}_{Faa})+{\cal G}_{Fab}{\cal
G}_{Fab}- {\cal G}_{Faa}{\cal G}_{Fbb}\over 
\bar T 
-{1\over 2}
{\cal C}_{ab}}
\biggr]\biggr\}
\label{Gamma2sp}
\end{eqnarray}
\noindent
The bosonic worldline Green's function ${\cal G}_B$ is
given in (\ref{calGBetc}), while the fermionic one is:
\begin{eqnarray}
{\cal G}_{F}(\tau_1,\tau_2)=G_{F12}\, {e^{-ieFT\dot{G}_{B12}} \over
\cos(eFT)}
\label{calGF}
\end{eqnarray}
where $G_{F12}={\rm sign}(\tau_1-\tau_2)$. 
For a self-dual background ${\cal G}_F$ simplifies to 
\begin{eqnarray}
{\cal G}_{F12}=G_{F12}\frac{\cosh(efT \dot{G}_{B12})}{\cosh(efT)}\Eins -i
G_{F12}\frac{\sinh(efT \dot{G}_{B12})}{efT\,\cosh(efT)}\,{\cal Z}
\label{sdgf}
\end{eqnarray}
The calculation, together with the charge and mass renormalization,
proceed in a manner analogous to the two-loop scalar QED calculation in
the previous section (for details of the mass renormalization, see
\cite{rss}). After similar manipulations we find that the on-shell
renormalized effective Lagrangian can be written as the following
proper-time integral:
\begin{eqnarray}
{\cal L}_{\rm spin}^{(2)(SD)}
&=&-\alpha \frac{m^4}{16\pi^3}\,\frac{2}{\kappa} \int_0^\infty dZ\, e^{-2\kappa Z} \left\{ \left(-Z +\coth Z \, \log\frac{\sinh Z}{Z}\right)\left(1-\frac{1}{4\kappa Z}\frac{d}{dZ}\right)\, \coth Z \right.
\nonumber\\
&&\left.+
\left(\log(2\kappa Z)+\gamma-\frac{5}{6}\right)\left(1-\frac{1}{8\kappa Z} \frac{d}{dZ}\right)\left(\coth^2 Z-\frac{1}{Z^2}-\frac{2}{3}\right)
+\frac{5}{6}+\frac{5}{24\kappa Z}\right\}
\non\\
\label{2lsppropertime}
\end{eqnarray}
Here we have also subtracted the zero-field term and done the charge renormalization. 

As in the scalar case, this proper-time integral can be done in
closed-form. After manipulations similar to those in the scalar case, 
we find the final result (\ref{2lspintro}),

\begin{eqnarray}
{\cal L}_{\rm spin}^{(2)(SD)}(\kappa)
=
-\alpha \,{m^4\over (4\pi)^3}\frac{2}{\kappa^2}\left[
3\xi^2 (\kappa)
-\xi'(\kappa)\right]
\nonumber
\end{eqnarray}
where $\xi(\kappa)$ is the same function that was defined before in
(\ref{defxi}). This result (\ref{2lspintro}) is remarkably similar to the
two-loop scalar QED result (\ref{2lscintro}). We find it interesting that the
spinor and scalar QED results for a constant selfdual background can each
be written in such a simple form in terms of the same function
$\xi(\kappa)$. This fact is discussed in the next section.

\section{Self-duality, helicity and supersymmetry at one- and two-loops}
\label{susy}
\renewcommand{\theequation}{4.\arabic{equation}}
\setcounter{equation}{0}

In this section we explain why the renormalized scalar and spinor effective 
Lagrangians in a self-dual background
are so similar, at both one- and two-loop. The self-duality of the background
has the effect that both the scalar and spinor cases are dramatically 
simplified from the results for a 
general constant background. But, on top of these simplifications, the scalar 
and spinor results (\ref{2lscintro}) and (\ref{2lspintro}) are remarkably 
similar to one another, each being expressed as a simple function
of $\xi(\kappa)$. The key to understanding this correspondence is the 
connection
between self-duality, helicity and supersymmetry.
It is well known that self-dual gauge fields 
are closely related to supersymmetry, and that they
correspond to helicity eigenstates
\cite{thooft,dadda,dufish,brolee}. 
In this section we explain how these connections are
manifest in our exact effective action results at both one- and two-loops.

Consider the one-loop QED case. Since $\coth^2 t=1+1/\sinh^2t$, we
compare the one-loop renormalized effective Lagrangians (\ref{1lsca}) and
(\ref{1lspa}) for the scalar and spinor cases, respectively, and find
that 
\begin{eqnarray}
{\cal L}_{\rm spin}^{(1)(SD)}(\kappa)=-2\, {\cal L}_{\rm
scal}^{(1)(SD)}(\kappa)
\label{susy1l}
\end{eqnarray}
Note that this proportionality between the scalar and spinor effective
Lagrangians is only true for a self-dual background. 
Furthermore, the relation
(\ref{susy1l}) holds for the renormalized effective Lagrangians, not
for the unrenormalized ones.

This proportionality between the one-loop scalar and spinor effective
Lagrangians reflects a well-known supersymmetry of the self-dual
background
\cite{thooft,dadda,dufish,brolee}. The basic relation between self-dual fields
and helicity can be traced to the following identity (we work in Euclidean space):
\begin{eqnarray}
\left(\gamma_\mu D_\mu
\right)^2\,\left(\frac{1+\gamma_5}{2}\right)=\left(D_\mu
D_\mu\right)\,\left(\frac{1+\gamma_5}{2}\right)
\label{sdh}
\end{eqnarray}
which holds provided the field strength $F_{\mu\nu}$ is self-dual. This
fact has the consequence \cite{thooft,dufish} that the bosonic and fermionic operators
\begin{eqnarray}
\Delta_B&=&- D_\mu D_\mu+m^2
\label{bop}\\
\Delta_F&=& -\left(\gamma_\mu D_\mu\right)^2 +m^2=
-D_\mu D_\mu-\frac{ie}{2}\sigma_{\mu\nu}F_{\mu\nu}+m^2
\label{fop}
\end{eqnarray}
have identical spectra, but the fermionic case has a four-fold multiplicity 
(from the spinor degrees of freedom). This is due to a quantum
mechanical supersymmetry \cite{dadda} relating 
the bosonic and fermionic operators in (\ref{bop}) and (\ref{fop}). 
Therefore,
\begin{eqnarray}
{\cal L}_{\rm spin}^{(1)(SD)}=\frac{1}{2}\log \det \Delta_F=2 \log \det \Delta_B =
-2 {\cal L}_{\rm scal}^{(1)(SD)}
\label{dets}
\end{eqnarray}
which is precisely the one-loop relation (\ref{susy1l}).

Another equivalent way to see this is to note that because of the
helicity/self-duality relation (\ref{sdh}), the spinor propagator in a
self-dual background can be expressed in a simple manner in terms of the
scalar propagator in the same background \cite{brolee}:
\begin{eqnarray}
S=- \left({D \hskip -7pt / -m}\right)\, G\,
\left(\frac{1+\gamma_5}{2}\right) - G\,{D \hskip -7pt /
}\,\left(\frac{1-\gamma_5}{2}\right)+\frac{1}{m}\left(1+ {D \hskip
-7pt /}\, G {D \hskip -7pt /}\right)\,\left(\frac{1-\gamma_5}{2}\right)
\label{sdprop}
\end{eqnarray}
where the scalar and spinor propagators are
\begin{eqnarray}
G=\frac{1}{-D^2+m^2} \qquad , \qquad S= \frac{1}{D \hskip -7pt / +m}
\label{props}
\end{eqnarray}
This relation between the spinor and scalar
propagators in a self-dual background provides another perspective on the
relation (\ref{susy1l}). The spinor induced current can be related to the
scalar induced current since, 
\begin{eqnarray}
\langle J^\mu_{\rm spin}\rangle
&=&
\frac{\delta}{\delta A_\mu} \, {\cal L}_{\rm spin}^{(1)}
\non\\
&=& -i e \tr \langle\gamma^\mu \, S\rangle\non\\
&=& 2 i e \langle D_\mu G + G D_\mu \rangle
\label{sdcurrents}
\end{eqnarray}
where in the last step we have used (\ref{sdprop}).
This current relation agrees  with the one-loop effective Lagrangian
relation (\ref{susy1l}). 

At the two-loop level, the supersymmetry relation (\ref{susy1l}) no longer
holds, as can be seen by comparing the two-loop scalar and spinor results
(\ref{2lscintro}) and (\ref{2lspintro}), respectively. Namely,
\begin{eqnarray}
{\cal L}_{\rm spin}^{(2)(SD)}(\kappa)
+2 {\cal L}_{\rm scal}^{(2)(SD)}(\kappa)
=-3 m^4 \frac{\alpha}{(4\pi)^3}
\, \frac{\xi^2(\kappa)}{\kappa^2}
\label{nosusy}
\end{eqnarray}
This non-vanishing is because the two-loop 
effective
Lagrangians are not simply logarithms of determinants.
Nevertheless, it is interesting to see
that each of the scalar and spinor two-loop effective Lagrangians can
be written in such a simple form in terms of the same function
$\xi(\kappa)$. To understand why this is the case, 
let us reconsider this computation in terms of 
more standard field theory techniques. 
Using Ritus's approach \cite{ginzburg}
the two-loop effective Lagrangians can 
be written as
\cite{ginzburg}:
\begin{eqnarray}
{\cal L}_{\rm spin}^{(2)}&=&\frac{e^2}{2} 
\int dx^\prime {\cal D}(x-x^\prime)\,
\tr\left[\gamma_\mu S(x,x^\prime)
\gamma_\mu S(x^\prime,x)\right]\label{2lspinor}\\
{\cal L}_{\rm scal}^{(2)}&=&-e^2\int dx^\prime {\cal D}(x-x^\prime)\, 
\left[D_\mu G(x,x^\prime)\,D_\mu G(x^\prime,x)
+D_\mu G(x,x^\prime) 
\mathrel{\mathop{D}^{\leftarrow}}_{\mu}
G(x^\prime,x)\right]
\label{2lscalar}
\end{eqnarray}
where $S(x,x^\prime)$ and $G(x,x^\prime)$ are the position space 
spinor and scalar propagators in the
presence of the background field, and ${\cal D}(x,x^\prime)$ denotes the 
internal photon propagator. We have not written tadpole terms, as these do 
not contribute after renormalization. 
These expressions (\ref{2lspinor}) and (\ref{2lscalar})
are valid for a  general background. Now, if we specialize to a self-dual background,
then the spinor and scalar propagators become related as in
(\ref{sdprop}), which means that the Dirac traces can be done in the
spinor case, leading to an expression in terms of the scalar propagator:
\begin{eqnarray}
{\cal L}_{\rm spin}^{(2)}=\frac{e^2}{2}\int dx^\prime {\cal D}(x-x^\prime)
\left[-8 D_\mu G(x,x^\prime)\,D_\mu G(x^\prime,x)
+16 D_\mu G(x,x^\prime) \mathrel{\mathop{D}^{\leftarrow}}_{\mu}
 G(x^\prime,x)
\right]
\label{2lspinorfin}
\end{eqnarray}
Thus, comparing with the scalar two-loop effective Lagrangian
(\ref{2lscalar}), we see that the spinor two-loop effective Lagrangian
for a self-dual background involves the same structures but with
different coefficients. This is reflected directly in our explicit
closed-form two-loop expressions (\ref{2lscintro}) and (\ref{2lspintro}).

To make this connection more precise, we consider a general linear
combination of these two structures:
\begin{eqnarray}
\int dx^\prime {\cal D}(x-x^\prime)
\left[A\, D_\mu G(x,x^\prime)\,D_\mu G(x^\prime,x)
+B\, D_\mu G(x,x^\prime) 
\mathrel{\mathop{D}^{\leftarrow}}_{\mu}
G(x^\prime,x)
\right]
\label{lincomb}
\end{eqnarray}
where $A$ and $B$ are numerical coefficients. Now, in a constant self-dual background,
the position space scalar propagator has the following simple form (up to a phase
that is not important here)
\begin{eqnarray}
G(x,x^\prime)=\frac{ef}{(4\pi)^2}\int_0^\infty \frac{dt}{\sinh^2(t)}\, \exp\left[ 
-2\kappa t-\frac{ef}{4}\coth(t) (x-x^\prime)^2\right]
\end{eqnarray}
This means that we can express the general linear combination in (\ref{lincomb})
as
\begin{eqnarray}
&&\frac{m^4}{(4\pi)^4 \kappa^2} \int_0^\infty dt \int_0^\infty dt^\prime 
\frac{e^{-2\kappa(t+t^\prime)}}{\sinh^2(t)\sinh^2(t^\prime)} 
\left\{ \frac{\frac{1}{2}(A+B)}{(\coth(t)+\coth(t^\prime))^2}
-\frac{\frac{A}{4}\coth(t+t^\prime)-\frac{B}{2}\kappa}{(\coth(t)+\coth(t^\prime))}
\right\}\non\\
&=&\frac{m^4}{(4\pi)^4 \kappa^2} \int_0^\infty Z dZ e^{-2\kappa Z} \int_0^1 du 
\left\{ \frac{\frac{1}{2}(A+B)}{\sinh^2(Z)}
-\frac{\frac{A}{2}\coth(Z)-B\kappa}{\sinh(Z)(\cosh(Z)-\cosh(Z(1-2u)))}
\right\}
\non\\
\label{newcomb}
\end{eqnarray}
where in the last step we have defined new integration variables $Z=t+t^\prime$, and
$u=\frac{t^\prime}{t+t^\prime}$. After renormalization, 
and using similar integration techniques as in section \ref{section2lsd},
the first term inside the
curly parentheses in (\ref{newcomb}) leads to $\xi^\prime$, with a numerical
coefficient $\frac{1}{2}(A+B)$, while the second term leads to $\xi^2$,
with a coefficient $-(\frac{A}{2}+B)$. Thus, the general linear
combination in (\ref{lincomb}) leads, after renormalization, to 
\begin{eqnarray}
&&\int dx^\prime {\cal D}(x-x^\prime)
\left[A\, D_\mu G(x,x^\prime)\,D_\mu G(x^\prime,x)
+B\, D_\mu G(x,x^\prime) 
\mathrel{\mathop{D}^{\leftarrow}}_{\mu}
G(x^\prime,x)
\right]\non\\
&=&\frac{m^4}{(4\pi)^4}\frac{1}{\kappa^2}\left[-(\frac{A}{2}+B)
\xi^2(\kappa)+
\frac{1}{2}(A+B) \xi^\prime(\kappa)\right]
\label{finalcomb}
\end{eqnarray}
Now, the two-loop scalar QED 
effective Lagrangian (\ref{2lscalar}) has numerical coefficients
$A=B=1$, and an overall factor of $-e^2$, which then leads immediately to
our scalar  result (\ref{2lscintro}). On the other hand, the two-loop
spinor QED  effective Lagrangian (\ref{2lspinorfin}) has
$A=-8$ and $B=16$, and an overall factor of $\frac{e^2}{2}$, which then
leads immediately to our spinor result (\ref{2lspintro}). Thus, the
simple relation between the two-loop spinor and scalar effective
Lagrangians for a self-dual background is ultimately due to the relation
(\ref{sdprop}) between the spinor and scalar propagators in a self-dual
background.

\section{The all `+' helicity amplitudes in the Euler-Heisenberg
approximation}
\label{amplitudes}
\renewcommand{\theequation}{5.\arabic{equation}}
\setcounter{equation}{0}

As is well known,
the Euler-Heisenberg Lagrangian contains the information about the
low-energy limit of the $N$ - photon amplitudes at the given loop level.
After specialization to a self-dual field it still contains
information about the component of this amplitude which has
all helicities alike.
This component of the $N$ - photon or gluon amplitudes 
generally leads to especially compact expressions 
\cite{ttwu,mahlon,bddgw,bgmv,mangano,bernreview,bededi,partay}.
In this section, we will
apply our results for the self-dual Euler-Heisenberg Lagrangians to
an explicit computation of these amplitudes in the low-energy limit,
at one and two loops.

At the one loop level it is not difficult to perform a direct
calculation of these amplitudes in the Bern-Kosower formalism.
We present this calculation, too, since it is quite instructive to
see how the equivalence between both approaches comes about.

The Bern-Kosower master formula \cite{berkos}, together with
a judiciously chosen set of partial integrations 
\cite{berkos,strassler2,menphoton}, allows one to write
the one-loop $N$ - photon amplitude in scalar QED in the following form
\footnote{In this section we use Minkowski space (+ - - -) conventions.}
:
\bear
\Gamma_{\rm scal}^{(1)}
[k_1,\varepsilon_1;\ldots ;k_N,\varepsilon_N]
&=&
-{e^N\over 
16\pi^2}
{\dps\int_{0}^{\infty}}{ds\over s}
e^{-im^2s}
s^{N-2}
\int_0^1du_1\int_0^1du_2\cdots\int_0^1du_N
\non\\
&&\times
Q_N(\dot G_{Bij})
\exp\biggl\lbrace -is\sum_{i<j=1}^N 
G_{Bij} k_i\cdot k_j
\biggr\rbrace
\label{Nphotonamplitude}
\ear\no
Here $G_{Bij} = \mid u_i-u_j\mid 
-{(u_i-u_j)}^2$, $\dot G_{Bij} = {\rm sign}(u_i-u_j)-2(u_i-u_j)$.
This representation holds on- and off-shell.
It is important to note that the right-hand side represents
the complete amplitude, not just a particular Feynman diagram.
See \cite{berdun,review}
for the relation of this representation
to standard Feynman parameter integrals. 
The exponential factor
$\exp\lbrace -is\sum_{i<j=1}^N G_{Bij} k_i\cdot k_j \rbrace$ 
corresponds to the standard one-loop $N$ - point denominator,
while $Q_N$ corresponds to a numerator polynomial. 
This polynomial depends on
the various $\dot G_{Bij}$'s 
as well as
on the kinematic invariants $\varepsilon_i\cdot k_j$, 
$\varepsilon_i\cdot\varepsilon_j$.
$Q_N$ can be decomposed in terms of `cycles' and `tails'
as follows
\footnote{We follow the notation of \cite{menphoton} rather than
\cite{review}.}: Define the `Lorentz cycle' $Z_n$ by
\bear
Z_2(ij)&\equiv&
\half {\rm tr}\Bigl(F_iF_j\Bigr)
\non\\
Z_n(i_1i_2\ldots i_n)&\equiv&
{\rm tr}
\Bigl(
F_{i_1}
F_{i_2}
\cdots 
F_{i_n}
\Bigr) 
\quad (n\geq 3)
\label{defZn}
\ear\no
where
$(F_i)_{\mu\nu} \equiv
k_{i\mu}\varepsilon_{i\nu}
- \varepsilon_{i\mu}k_{i\nu}$
is the field strength tensor for leg `i'.
Define 
the `bi-cycle of length $n$', $\dot G (i_1i_2\ldots i_n)$, 
by 
\bear
\dot G (i_1i_2\ldots i_n)
&\equiv& 
\dot G_{Bi_1i_2} 
\dot G_{Bi_2i_3} 
\cdots
\dot G_{Bi_{n-1}i_n}
\dot G_{Bi_ni_1}
Z_n(i_1i_2\ldots i_n)
\label{defbicycle}
\ear
Then $Q_N$ has a decomposition of the form
\bear
Q_N  &=& \sum Q_N^{z_1z_2\cdots z_i}
\label{cycledecomp}
\ear
where the superscripts on a term indicate that it contains
$i$ bi-cycle factors of lengths $z_1,\ldots,z_i$. Unless all
indices $1,\ldots ,N$ in a given term are bound up in cycle factors 
there will be one more factor, called `tail'. 

One advantage of this representation is that the polynomial 
$Q_N$ is homogeneous in the momenta; each term in it has 
exactly $N$ factors of momentum. Therefore the low-energy limit
of the amplitude, corresponding to the case where $m^2$ 
becomes much larger than all kinematic invariants
$k_i\cdot k_j$, is simply obtained by replacing the 
exponential factor 
$\exp\lbrace -is\sum_{i<j=1}^N 
G_{Bij} k_i\cdot k_j
\rbrace$
by unity. The omission of this factor not only decouples the
global proper-time integral, but also turns all terms involving tails
into total derivatives. This leaves us with only those numerator 
terms which are products of bi-cycles. Thus the 
low energy, or Euler-Heisenberg (`EH'),
limit of this amplitude can be written as
\bear
\Gamma_{\rm scal}^{(1)(EH)}
[k_1,\varepsilon_1;\ldots ;k_N,\varepsilon_N]
&=&
{(-ie)^N\Gamma(N-2)\over 
16\pi^2m^{2N-4}}
\int_0^1du_1\int_0^1du_2\cdots\int_0^1du_N
\,Q_N^{(EH)}(\dot G_{Bij})
\non\\
\label{Nphotonlimit}
\ear\no
where 
\bear
Q_N^{(EH)}(\dot G_{Bij})
&=&
\sum_{\rm part.}
P(\lbrace  i_1i_2\ldots i_{k_1}\rbrace)
P(\lbrace i_{k_1+1}\ldots i_{k_1+k_2}\rbrace)
\cdots
P(\lbrace i_{k_1+k_2+\ldots k_{j-1}+1}\ldots i_N\rbrace)
\non\\
\label{defQNEH}
\ear
Here the sum is over all possible partitions
of the set of indices ${1,\ldots,N}$
into subsets, and $P(\lbrace i_1i_2\ldots i_k\rbrace)$
denotes the sum over all {\sl distinct}
bi-cycles which can be formed with the given
subset of indices, e.g. 
$P(\lbrace  ijkl\rbrace) = \dot G(ijkl) + \dot G(ijlk) + \dot G(ikjl)$.
The integral $\int_0^1 du_1\cdots du_N$ 
factorizes into copies of the
basic `chain integral' \cite{ss1}
\bear
b_n &\equiv& \int_0^1 du_1du_2\ldots du_n\,
\dot G_{B12}\dot G_{B23}\cdots\dot G_{Bn1} =
\qquad\left\{ \begin{array}{r@{\quad\quad}l}
-2^n{{\cal B}_n\over n!}  & \qquad n{\rm \quad even}\\
0 & \qquad n{\rm \quad odd}\\
\end{array} \right.
\label{chainintbos}
\ear
where ${\cal B}_n$ denotes the $n$th Bernoulli number.
The result of the integrations can therefore be written as
\bear
\int_0^1du_1\int_0^1du_2\cdots\int_0^1du_N
\,Q_N^{(EH)}(\dot G_{Bij})
&&=
\non\\&&
\hspace{-160pt}
\sum_{\rm part.}
b_{k_1}F_{k_1}(\lbrace i_1i_2\ldots i_{k_1}\rbrace)
b_{k_2}F_{k_2}(\lbrace i_{k_1+1}\ldots i_{k_1+k_2}\rbrace)
\cdots
b_{k_j}F_{k_j}(\lbrace i_{k_1+k_2+\ldots k_{j-1}+1}\ldots i_N\rbrace)
\non\\
\label{intu}
\ear
where 
$F_k(\lbrace i_1i_2\ldots i_k\rbrace)$
denotes the sum over all distinct
Lorentz cycles which can be formed with a given
subset of indices, e.g. 
$F_4(\lbrace ijkl\rbrace) = Z_4(ijkl) + Z_4(ijlk) + Z_4(ikjl)$.
This allows us to write the amplitude in the
following form:
\bear
\Gamma_{\rm scal}^{(1)(EH)}
[k_1,\varepsilon_1;\ldots ;k_N,\varepsilon_N]
&=&
{(-ie)^N (N-1)!\over 
16\pi^2m^{2N-4}}
\,\exp\biggl\lbrace \sum_{m=1}^{\infty}b_{2m}
\sum_{\lbrace i_1\ldots i_{2m}\rbrace}
F_{2m}(\lbrace i_1i_2\ldots i_{2m}\rbrace)\biggr\rbrace
\bigg \vert_{(1\ldots N)}
\non\\
\label{Nphotonlimitfin}
\ear\no
Here the second sum in the exponent runs over 
the set of all sets of $2m$ different positive numbers, and it is understood
that after expansion of the exponential only those terms are kept which
contain each index $1,2,\ldots,N$ precisely once. 

It takes only a minor modification to generalize 
this computation to the spinor loop case: According to
the Bern-Kosower `replacement' rules, the transition from 
scalar to spinor QED can be effected by replacing the
above `chain integral' (\ref{chainintbos}) by the
`super chain integral' 
\bear
b_n-f_n &\equiv& \int_0^1 du_1du_2\ldots du_n\,
\Bigl(\dot G_{B12}\dot G_{B23}\cdots\dot G_{Bn1} 
-
G_{F12}G_{F23}\cdots G_{Fn1}\Bigr)
=
(2-2^n)\,b_n \non\\
\label{chainintferm}
\ear
As usual one must also multiply by a global factor
of $(-2)$ for statistics and degrees of freedom.

Let us now compare with the calculation of the same amplitude
via the effective action. In (\ref{L1scalren}) we wrote the
one-loop Euler-Heisenberg Lagrangian in terms of the two Maxwell
invariants. Alternatively, the integrand can also be written in terms
of traces of powers of the field strength tensor \cite{rss}:
\bear
{\cal L}^{(1)}_{\rm scal}(F) &=&
-{1\over 16\pi^2}\int_0^{\infty}{ds\over s^3}
\,\e^{-ism^2}
{\rm exp}\biggl\lbrace
\frac{1}{2}\sum_{m=1}^{\infty}
{b_{2m}\over 2m}(es)^{2m}
{\rm tr}[F^{2m}]
\biggr\rbrace
\label{L1scalF}
\ear
According to standard field theory (see, e.g., \cite{itzzubbook})
the low energy limit of the $N$ - photon
amplitude can be obtained from the effective
Lagrangian by replacing 
$F$ with $i\sum_{i=1}^N F_i$, expanding out, and collecting all
those terms which involve each $F_1,\ldots,F_N$, precisely
once. The equivalence of (\ref{Nphotonlimitfin}) and (\ref{L1scalF})
therefore follows immediately from the combinatorial fact that
\bear
{\rm tr}\Bigl[(F_1+\ldots +F_N)^n\Bigr]\bigg\vert_{\rm all\,\,different}
&=&
2n \sum_{\lbrace i_1\ldots i_n\rbrace} F_n(\lbrace i_1i_2\ldots i_n\rbrace)
\qquad (2\le n \le N)
\non\\
\label{F=F}
\ear
Here the sum runs over all possiblities of choosing a
subset of $n$ indices out of $\lbrace 1,\ldots,N\rbrace$.
On the left hand side it is understood that terms containing
the same $F_i$ twice are to be discarded.

To obtain the on-shell helicity amplitudes, one now needs to
construct polarisation vectors for all possible
inequivalent helicity assignments, and compute the various
$\varepsilon_i\cdot k_j$'s and $\varepsilon_i\cdot\varepsilon_j$'s.
For small numbers of photons, this could be easily worked out
for arbitrary assignments, using standard techniques such as the
spinor helicity formalism \cite{bkdgw,klesti,xuzhch,dixon,ttwu}.
In this formalism, a polarisation vector with circular polarisation
`$\pm$' for a photon with momentum $k$ is written as
\bear
\varepsilon^{\pm}_{\mu} &=&
\pm
{\langle q^{\mp}\mid\gamma_{\mu}\mid k^{\mp}\rangle
\over
\sqrt{2}\langle q^{\mp}\mid k^{\pm}\rangle}
\label{eps+-}
\ear
Here 
$\langle q^{\pm}\mid k^{\mp}\rangle = \overline{u_{\pm}(q)}u_{\mp}(k)$
etc. are basic spinor products, and $q$ is a reference momentum which 
is different from $k$ but arbitrary otherwise; see \cite{dixon} for
details and conventions. 
We will be concerned here only with the all `+' amplitudes
(see, however, the last remark at the end of this section). 
For those, we know that all field strength tensors
$F_1,\ldots,F_N$, and therefore also their sum
$F_{\rm tot}\equiv \sum_{i=1}^NF_i$, are self-dual
tensors. Therefore, as in (\ref{idF2})
the square of $F_{\rm tot}$ must be proportional to
the Lorentz identity. And indeed, using the spinor helicity
formalism it is easy to see that
\bear
F_i^2 &=& 0 \non\\
(F_iF_j + F_jF_i)^{\mu\nu} &=& -\half [ij]^2\eta^{\mu\nu} \non\\
\label{Fids}
\ear
where $[i\,j]\equiv \langle {k_i}^+\mid {k_j}^-\rangle$.
It follows that
\bear
F_{\rm tot}^2 &=& -f_{\rm tot}^2\Eins \non\\
f_{\rm tot}^2 &\equiv & \half\sum_{1\le i < j \le N}[ij]^2 \non\\  
\label{Ftotsquare}
\ear
and
\bear
{\rm tr}[F_{\rm tot}^{2m}] &=& 4(-1)^m\Bigl( f_{\rm tot}^2\Bigr)^m
\label{Ftottoftot}
\ear
Thus, the all `+' component of the
$N$ - point amplitude can be obtained from the
self-dual effective Lagrangian (\ref{1lsca}) as follows:
Expand this Lagrangian in powers of $f$, i.e. in the weak field
expansion; explicitly,
\begin{eqnarray}
{\cal L}_{\rm scal}^{(1)(SD)}(f)
&=&\frac{m^4}{(4\pi)^2}\,\sum_{n=2}^\infty c_{\rm scal}^{(1)}(n)
{(2ef)^{2n}\over m^{4n}} \non\\
c_{\rm scal}^{(1)}(n)&=& - \frac{{\cal B}_{2n}}{2n(2n-2)}
\label{1lscweak}
\end{eqnarray}
The $N=2n$  point amplitude is then obtained 
simply by taking the term involving $f^N$, replacing 
$f^2$ by $f_{\rm tot}^2$, and keeping only those terms in
the expansion of $f_{\rm tot}^N$ involving each index precisely
once:
\bear
\Gamma_{\rm scal}^{(1)(EH)}
[k_1,\varepsilon_1^+;\ldots ;k_N,\varepsilon_N^+]
&=&
\frac{(2e)^{N}}{(4\pi)^2m^{2N-4}}\,c_{\rm scal}^{(1)}
({\scriptstyle\frac{N}{2}})
\chi_N \non\\
\chi_N &\equiv & f_{\rm tot}^N\Big\vert_{\rm all\,\, different}
\non\\
&=& 
{({\scriptstyle \frac{N}{2}})!
\over 2^{N\over 2}}
\Bigl\lbrace
[12]^2[34]^2\cdots [(N-1)N]^2 + {\rm \,\, all \,\, permutations}
\Bigr\rbrace
\non\\
\label{ampscal1}
\ear
To generalize this result to the spinor QED and two-loop cases,
all we need are the corresponding weak field expansions. For
the one-loop amplitudes in spinor QED, we infer from (\ref{susy1l})
that they are proportional to the scalar QED ones:
\bear
\Gamma_{\rm spin}^{(1)(EH)}
[k_1,\varepsilon_1^+;\ldots ;k_N,\varepsilon_N^+]
&=&
-2\Gamma_{\rm scal}^{(1)(EH)}
[k_1,\varepsilon_1^+;\ldots ;k_N,\varepsilon_N^+]
\label{ampsusy}
\ear
At the two-loop level, we can easily obtain the weak field
expansions of the Lagrangians (\ref{2lscintro}), (\ref{2lspintro})
from the well-known asymptotic expansion of the
digamma function at infinity (see part 2). This yields
\bear
\Gamma_{\rm scal}^{(2)(EH)}
[k_1,\varepsilon_1^+;\ldots ;k_N,\varepsilon_N^+]
&=&
\alpha\pi\frac{(2e)^{N}}{(4\pi)^2m^{2N-4}}\,c_{\rm scal}^{(2)}
({\scriptstyle{\frac{N}{2}}})
\chi_N 
\non\\
c^{(2)}_{\rm scal}(n)&=&
{1\over (2\pi)^2}\biggl\lbrace
\frac{2n-3}{2n-2}\,{\cal B}_{2n-2}
+\frac{3}{2}\sum_{k=1}^{n-1}
{{\cal B}_{2k}\over 2k}
{{\cal B}_{2n-2k}\over (2n-2k)}
\biggr\rbrace
\non\\
\label{ampscal2}
\ear
for scalar QED, and
\bear
\Gamma_{\rm spin}^{(2)(EH)}
[k_1,\varepsilon_1^+;\ldots ;k_N,\varepsilon_N^+]
&=&
-2\alpha\pi\frac{(2e)^{N}}{(4\pi)^2m^{2N-4}}\,c_{\rm spin}^{(2)}
({\scriptstyle{\frac{N}{2}}})
\chi_N 
\non\\
c^{(2)}_{\rm spin}(n) &=&
{1\over (2\pi)^2}\biggl\lbrace
\frac{2n-3}{2n-2}\,{\cal B}_{2n-2}
+3\sum_{k=1}^{n-1}
{{\cal B}_{2k}\over 2k}
{{\cal B}_{2n-2k}\over (2n-2k)}
\biggr\rbrace
\non\\
\label{ampspinl2}
\ear
for spinor QED.

Let us discuss some implications of these results.
In
the massless case the one-loop all `+' $N$ - photon amplitudes 
have been shown to vanish for $N\ge 6$ \cite{mahlon}. Our result shows
that this miracle does not generalize to the massive case at either one 
or two loops.

The one-loop relation (\ref{ampsusy}) is a direct consequence of
the supersymmetry Ward identities for S-matrix elements first obtained
by Grisaru, Pendleton and van Nieuwenhuizen \cite{grpeva,gripen}.
Those identities hold for massive as well as for massless supermultiplets in
the loop.  
As was already clear from (\ref{nosusy}) this relation (\ref{ampsusy}) does
not generalize to the two-loop level. The reason is, of course,
that at two loops ordinary QED `knows' already that it is not
supersymmetric. The corresponding supersymmetry identity at the two-loop
level involves, in addition to the scalar and fermion loop pieces
(\ref{ampscal2}), (\ref{ampspinl2}), contributions from diagrams with
a photino exchange as well as diagrams containing the scalar quartic coupling.
At two loops our results 
could still be used together with the supersymmetry
Ward identity to infer the sum of those additional contributions 
to the `all +' amplitude in the Euler-Heisenberg approximation.
See the recent \cite{bgmv} for an explicit verification of the
vanishing of the two -- loop four -- photon `all +' amplitude
in massless $N=1$ and $N=2$ SUSY QED.   

Now that we have the above explicit all - $N$ formulas, it would be
of obvious interest to study
the large $N$ behaviour of the coefficients 
$c^{(1,2)}_{\rm scal,spin}({\scriptstyle \frac{N}{2}})$.
This will be done in detail in part 2. For now,
let us just anticipate the facts that the leading asymptotic growth
of the coefficients turns out to be factorial, and 
to be the same at one and two loops.

Finally, let us mention that the same method of computation
applied to the photon amplitudes with all but one helicity alike 
$(-++\ldots +)$ allows us to show without computation that
those vanish in the low energy limit. 
Namely, gauge invariance implies that the Euler-Heisenberg
Lagrangian can, at arbitrary loop order and at an arbitrary
fixed order in the weak field expansion, be written as
a linear combination of terms involving only products
of ${\rm tr}[F^n]$'s. After substituting for $F$ the sum
of the individual $F_i$'s this yields sums of terms
involving only products of $Z_n$'s, where one of the
factors must contain the $F_i$ which has negative helicity,
say, $F_1$. But using the spinor helicity formalism it is
easy to show that a $\tr (F_1^-F_{i_1}^+\cdots F_{i_k}^+)$
always vanishes (use $k_1$ as the reference momentum for
all other legs). 
Therefore the Euler-Heisenberg limits of these amplitudes
must vanish to all orders
in perturbation theory. This argument applies, of course, to both
scalar and spinor QED.

\section{Conclusions}
\label{conclusions}
\renewcommand{\theequation}{6.\arabic{equation}}
\setcounter{equation}{0}

To summarize, we have computed the on-shell renormalized two-loop
effective
Lagrangian for both scalar and spinor QED, for the case of a constant
self-dual Euclidean background. This case exhibits remarkable simplifications
compared to the general constant background, and even compared to the constant
magnetic (or electric) background. For the self-dual case the propagators 
simplify to such an extent that the final renormalized answers can be given in
a simple closed form in terms of a common function $\xi(\kappa)$, which is
essentially the digamma function. 
Although for $\phi^4$ theory an analogous result
has been known for a long time \cite{ilitma}
we are not aware of any comparable
two-loop result in gauge theory.

Furthermore, the scalar and spinor QED cases
are very similar to one another, a fact that we explained using the relation 
between
self-duality, supersymmetry and helicity, which has the implication that 
the spinor propagator in a self-dual background has a simple expression in 
terms of the scalar propagator in the same background. At one-loop this
implies that the scalar and spinor effective Lagrangians are proportional, but
at two-loop the relation is more involved. This provides a new
manifestation of the well-known simplicities arising in helicity 
(or self-duality)
projections. 

Our explicit results for the self-dual two-loop effective Lagrangians 
have allowed us to compute the low energy limit of the
massive $N$ - photon amplitudes with all helicities alike, 
at one and two loops, for both scalar and spinor QED. This shows,
in particular, that the vanishing theorem mentioned above
does not extend to massive amplitudes. 
As a side result, we have shown that the amplitudes with
all but one helicity alike vanish in the same approximation,
to all orders in perturbation theory.

From the mathematical point of view, we find it intriguing that,
for both scalar and spinor QED and at both one and two loops,
the final results for the effective Lagrangian 
turned out to be expressible in terms of the
`mother' function $\Xi(\kappa )$ and its derivatives. This property
might be accidental, but we have reasons to suspect that,
at least for QED in the quenched approximation,
it will be found to extend to arbitrary loop orders. 

Moreover, we have seen that 
even in QCD the same would hold true at one loop if a gluon
mass was introduced (say, as an infrared regulator). 
It would be very interesting to know the form of the result for
the nonabelian case at two loops, i.e. the two loop effective
Lagrangian for Yang-Mills theory in a constant self-dual quasi-abelian
background. The two-loop worldline formulas derived in
\cite{sascza} for the general nonabelian Euler-Heisenberg type
Lagrangian should provide a good starting point for its
computation.

Our results also suggest a new approach to the 
computation of the Lagrangians ${\cal L}^{(2)}_{\rm scal,spin}$
for a general constant background. Rather than performing a
double expansion in the variables $a,b$, as has been done 
in \cite{ginzburg,korsch}, it might be preferable to use an
expansion in the variables $H\equiv a+ib$ and $\bar H\equiv a-ib$.
The integration techniques which we have used here for the
self-dual case, corresponding to $\bar H = 0$, would 
in principle also allow
one to obtain, at any fixed order of this expansion in ${\bar H}^n$,
a closed form expression for the coefficient as a function of
$H$. This might lead to progress even in the computation of the
magnetic and electric cases.

We anticipate that the
techniques developed here may also help to extend the 
one-loop instanton background calculations \cite{thooft,carlitz,kwon} to 
the two-loop level.

In part 2 of this paper \cite{sdtwo}, we investigate the structure of the 
imaginary part of the two-loop self-dual
Euler-Heisenberg Lagrangians for the `electric' case.
We also study the weak- and strong-field expansions of
these two-loop effective Lagrangians, and use them to test the techniques
of Borel summation, and the associated analytic continuation properties.

\vspace{20pt}
\noindent
{\bf Acknowledgements:} 
We thank Zvi Bern and Gregory Mahlon for correspondence concerning 
helicity structures in massive loop amplitudes.
Discussions and correspondence with Vladimir 
Ritus and Arkady Tseytlin are gratefully acknowledged. 
Both authors thank Dirk Kreimer and the Center for Mathematical Physics,
Mathematics Department, Boston University, for hospitality during the
final stage of this work.
We gratefully acknowledge the support of NSF and CONACyT
through a US-Mexico collaborative research grant, NSF-INT-0122615.

\end{document}